\def\be {\begin{equation}}
\def\ee {\end{equation}}
\def\bea {\begin{eqnarray}}
\def\eea {\end{eqnarray}}
\def\bc {\begin{center}}
\def\ec {\end{center}}
\def\bfg {\begin{figure}}
\def\efg {\end{figure}}
\def\bi {\begin{itemize}}
\def\ei {\end{itemize}}

\def\le {\left}
\def\ri {\right}
\def\pa {\partial}

\def\a  {\alpha}
\def\b  {\beta}
\def\c  {\gamma}

\def\d  {\delta}
\def\D  {\Delta}

\def\f {\phi}

\def\r  {\rho}

\def\t  {\tau}

\def\hb {\hbar}

\documentclass[onecolumn,preprintnumbers,amsmath,amssymb]{revtex4}
\usepackage{graphicx}

\begin{document}
\title{Quantum Gravity Corrections in Chandrasekhar Limits}
\author{Mohamed Moussa}
\email{mohamed.ibrahim@fsc.bu.edu.eg}
\email{moussa.km@gmail.com}
\affiliation{Physics Department, Faculty of Science, Benha University, Benha 13518, Egypt}

\begin{abstract}
It is agreed that Chandrasekhar mass and central density of white dwarfs are independent, which means that there is a whole series of stars having radius and central density as parameters that all have the same Chandrasekhar mass. In this article the influence of a quantum gravity is shown so the Chandrasekhar limits (mass and radius) depend explicitly on the central density and gravity parameters. A new polytropic relation between degenerate pressure of the star and its density is investigated. This leads to a modification in Lane-Emden equation and mass and radius formulas of the star. A modified Lane-Emden equation is solved numerically with consideration to the mass density of the star depends on its radius. The solution used in calculating the mass and radius limit of the white dwarf. It was found that mass and radius limits decrease due to increase in central density and gravity parameters in a comparison with the original values. We can say that central density and quantum gravity constitute a new tool that can help to make the theoretical values corresponding to experimental observations apply in a better manner.
\end{abstract}
\maketitle
\section{Introduction}
The most important prediction of a various quantum theories of gravity is the existence of minimum measurable length and a maximum measurable momentum near the Planck scale, and hence the continuum picture of spacetime breaks down. The black holes physics indicates that the minimum length is of order of the Planck length which should act as an universal feature of all models of quantum gravity  Ref.\cite{1,2}. Also, in the context of perturbative string theory, the strings are the smallest probe available, and so, it is not possible to probe the spacetime below the string length scale. Thus the string length scale acts as a minimum measurable length in string theory Ref.\cite{3,4,5,6,7}. This makes generalization process of uncertainty principle is a significant request in physics. The results is a presence of a deformed or generalized  uncertainty principle (GUP) which is equivalent to a modification in the commutation relations between position coordinates and momenta (deformed Heisenberg algebra) Ref.\cite{6,7,8,9,10,11,12,13,14,15}. It is known that the familiar uncertainty relation is closely related to the canonical Heisenberg algebra, this way the modified canonical Heisenberg algebra is related to a non-canonical one. So that the commutator of the position and momentum operators becomes momentum dependent, instead of a constant. With this non-canonical algebra, the coordinate representation of the momentum operator get modified, and this in turn produces correction terms for all quantum mechanical systems. On the other hand, Doubly Special Relativity (DSR), another approach to quantum gravity, leads to a deformation in Heisenberg algebra,\cite{16,17,18}. It inspires both the velocity of light and Planck energy as universal constants. The deformed Heisenberg algebra studied in DSR theory has been predicted from many consequences, such as, discrete spacetime Ref.\cite{19}, spontaneous symmetry breaking of Lorentz invariance in string field theory Ref.\cite{18}, spacetime foam models Ref.\cite{20}, spin-network in loop quantum gravity Ref.\cite{21}, non-commutative geometry Ref.\cite{22}, and Ho\^{r}ava-Lifshtz gravity Ref.\cite{23}. Another approach to quantum gravity through a modification in dispersion relation, this condition implies a breakdown of Lorentz symmetry Ref.\cite{20,123,223}. This model was tested extensively in physics, for example Ref.\cite{323,423,523,623}.

Ali, Das and Vagenas Ref.\cite{24,25} worked on a new approach for quantum gravity. They suggested commutators that are consistent with string theory, black hole physics and DSR and ensure $\le[x_i,x_j\ri]=\le[p_i,p_j\ri]=0$ (via Jacobi identity). The new commutator have the following form
\begin{equation}\label{1}
\le[x_i,p_j\ri]=i\hb \le[ \d_{ij}-\a \le( p\a_{ij}+\dfrac{p_ip_j}{p} \ri)+\a^2(p^2\d_{ij} + 3p_ip_j ) \ri]
\end{equation}
where $\a=\frac{\a_0}{M_{P}c}=\frac{\a_0l_{P}}{\hb}$. Where $M_{P}$, $l_{P}$ and $M_{P}c^2$ are Planck mass, length and energy, respectively. This in turn imply a minimum measurable length and a maximum measurable momentum in such a way $\D x\geq (\D x)_{min}\approx \a_0~l_{P}$ and $\D p\leq (\D p)_{max}\approx \frac{M_{P}~c}{\a_0}$. As a result, according to Eq.(\ref{1}), GUP modifies the physical momentum Ref.\cite{24,25,26}
\begin{equation}\label{3}
x_i=x_{0i}~~~~,~~~~p_i=p_{0i}\le(1-\a p_0+2\a^2p_0^2\ri)
\end{equation}
where $x_{0i}$ and $p_{0i}$ satisfying the canonical commutation relation $\le[x_{0i},p_{0i}\ri]=i\hbar \d_{ij}$, so we can consider $p_i$ as a momentum in Planck scale and $p_{0i}$ as a momentum at low energies (having standard representation in position space $p_{0i}=-i\hbar\frac{\pa}{\pa x_{0i}}$). It is assumed that the dimensionless parameter $\a_0$ is of the order of unity, in which case the $\a$ dependent terms are important when energies (momenta) are comparable to Planck energy (momenta) and length is comparable to the Planck length Ref.\cite{24}. The upper bounds on the GUP parameter $\a$ have been derived in a lot of previous works for example Refs.\cite{26,27,28,29,30,31,32}.

On the other hand the current observation indicates that the white dwarf has smaller radius than the theoretical predictions. This is lead us to introduce quantum gravity as a tool to explain this defect. It is worth mentioning that this problem is considered with many quantum gravity approaches. In Ref. \cite{34} Camacho assume that a constant density of white dwarf and he calculated the star radius with the modified dispersion relation that caused by a breakdown in Lorentz symmetry. It is found that the change in the star radius depends on the sign of the quantum gravity parameter. Amelino-Camélia et al Ref. \cite{35} try to improve these results using a small modification in density of state by assuming that the law of composition of momenta affects the rules of integration over energy-momentum space and these are crucially relevant for Chandrasekhar analysis. The authors in \cite{36}  extend the analysis of Camacho by stopping the unphysical assumption of constant density. They assumed that  the density is not constant throughout the star and reported the numerical solutions to the exact equations for the Chandrasekhar. The result is  the realistic density shows a significant corrections at Planck scale and the mass limit is raised or lowered according to the sign of the modification. The stability of white dwarfs is also examined using a non-commutative geometry concept in \cite{37}. Another approach of generalized uncertainty principle is contracted, concepts of this approach is reported Ref \cite{7,14,15,38}. Using this approach the author in \cite{39,40,140} found that  quantum gravity correction depends on the number density of the star and it tends to resist the star collapse.

We will consider Chandrasekhar limit with this quantum gravity approach. The analysis in \cite{41,42} will be extended by disregarding the unphysical assumption of constant density. In fact pressure and density of the star depend on the star radius. The modified star pressure will be calculated which leads to a modification in Lane-Emden equation. This equation will be solved numerically in order to determine the the quantum gravity corrections on the mass and radius of the star.

\section{Modified Pressure inside White Dwarfs}

In order to investigate quantum gravity effects the statistical mechanical equations should be put in a form consistent with GUP framework. The GUP can be considered in phase space analysis by two equivalent ways. First considering deformed commutator with non-deformed Hamiltonian function (i.e. deformed the measure of integration). Second calculating canonical variables on the GUP corrected phase space which satisfy the standard commutative algebra (i.e. non-deformed standard measure of integration), in this case the Hamiltonian function should be deformed. These two pictures are related to each other by Darboux theorem. In this work we will consider a deformed measure of integration with non-deformed Hamiltonian function. In order to do that the effect of GUP on the density of states should be considered in such that the number of microstates inside the phase space volume does not change with time, this what is called Liouville theorem  \cite{33,333}. To do that let us begin with the Heisenberg equations of motion
\begin{eqnarray}\label{v1}
\dot{x}=\dfrac{1}{i\hbar}\le[x,H\ri]~~~~,~~~~\dot{p}=\dfrac{1}{i\hbar}\le[p,H\ri]
\end{eqnarray}
Using the correspondence principle between commutator in quantum mechanics and poisson bracket in classical mechanics the classical limit of the equation of motions are
\begin{eqnarray}
\dot{x_i}=\{x_i,H\}=\{x_i,p_j\}~\dfrac{\pa H}{\pa p_j} \label{v2}\\
\dot{p_i}=\{p_i,H\}=-\{x_j,p_i\}~\dfrac{\pa H}{\pa x_j} \label{v3}
\end{eqnarray}
Now suppose the evolution of coordinate and momentum  during an infinitesimal time interval $\d t$ such that
\begin{eqnarray}
x_i'=x_i+\d x_i  \label{v4} \\
p_i'=p_i+\d p_i  \label{v5}
\end{eqnarray}
Taking into account Eqs. (\ref{v2},\ref{v3}) the infinitesimal change in coordinate and momentum may be written as
\begin{eqnarray}
\d x_i=\{x_i,p_j\}~\dfrac{\pa H}{\pa p_j}~\d t \label{v6} \\
\d p_i=-\{x_j,p_i\}~\dfrac{\pa H}{\pa x_j}~\d t  \label{v7}
\end{eqnarray}
After infinitesimal evolution the infinitesimal phase space volume should be
\begin{eqnarray}
\nonumber d^Dx'~d^Dp'=\le|\dfrac{\pa(x'_1,...,x'_D,p'_1,...,p'_D)}{\pa(x_1,...,x_D,p_1,...,p_D)}\ri|d^Dx~d^Dp~~~~\\
=\le[1+\le(\dfrac{\pa \d x_i}{\pa x_i}+\dfrac{\pa \d p_i}{\pa p_i}\ri)+...\ri]d^Dx~d^Dp  \label{v8}
\end{eqnarray}
In Eq.(\ref{v8}) we kept the Jacobian terms of first order in $\d t$ and used Eqs.(\ref{v4},\ref{v5}). Now by using Eqs.(\ref{v6},\ref{v7})  we can prove that
\begin{eqnarray}
\nonumber \le(\dfrac{\pa \d x_i}{\pa x_i}+\dfrac{\pa \d p_i}{\pa p_i}\ri)\dfrac{1}{\d t}~~~~~~~~~~~~~~~~~~~~~~~~~~~~~~~~ \\
\nonumber =\dfrac{\pa }{\pa x_i} \le[\{x_i,p_j\}\dfrac{\pa H}{\pa p_j}\ri]-\dfrac{\pa }{\pa p_i}\le[\{x_j,p_i\}\dfrac{\pa H}{\pa x_j}\ri]\\
\nonumber =-\dfrac{\pa}{\pa p_i}\le[\d_{ij}-\a \le(p\d_{ij}+\dfrac{p_ip_j}{p}\ri)\ri]~~~~~~~~~~~~\\
=\a (D+1)\dfrac{p_j}{p}\dfrac{\pa H}{\pa x_j}~~~~~~~~~~~~~~~~~~~~~~~~~~~~~~~\label{v9}
\end{eqnarray}
Use Eq.(\ref{v9}) into Eq.(\ref{v8})
\begin{eqnarray}
\nonumber d^Dx'~d^Dp'=\le[1+\a (D+1)\dfrac{p_j}{p}\dfrac{\pa H}{\pa x_j}~\d t\ri]d^Dx~d^Dp \\
=\le[1+\a \dfrac{p_j}{p}\dfrac{\pa H}{\pa x_j}~\d t\ri]^{D+1}d^Dx~d^Dp ~~~~ \label{v10}
\end{eqnarray}
Also we can prove that
\begin{eqnarray}
\nonumber 1-\a p'=1-\a \sqrt{p_i'p_i'}~~~~~~~~~~~~~~~~~~~~~\\
=1-\a \sqrt{(p_i+\d p_i)(p_i+\d p_i)} \label{v11}
\end{eqnarray}
keeping only terms proportional to $\a$ and use Eq.(\ref{v7}), one gets
\begin{eqnarray}
1-\a p'\simeq(1-\a p)\le[1+\a \dfrac{p_j}{p}\dfrac{\pa H}{\pa x_j}~\d t\ri] \label{v12}
\end{eqnarray}
Use Eq.(\ref{v12}) into Eq.(\ref{v10}) and integrate over coordinate space volume
\begin{eqnarray}
\dfrac{V}{h^D}\dfrac{d^Dp'}{(1-\a p')^{D+1}}=\dfrac{V}{h^D}\dfrac{d^Dp}{(1-\a p)^{D+1}} \label{v13}
\end{eqnarray}
This equation shows that the number of states inside a volume of phase space does not change with time evolution in the GUP regime. So it is clear that the density of microstates should be modified by the factor $(1-\a p)^{-D-1}$. According to these concepts the modification in the number of quantum states per momentum per space volume, in $3D$ dimensions, should be
\begin{eqnarray}
\dfrac{4\pi V}{h^3}\int p^2dp \rightarrow \dfrac{4\pi V}{h^3}\int \dfrac{p^2dp}{(1-\a p)^4} \label{v14}
\end{eqnarray}

On the other hand a white dwarf star is the end product of stellar evolution, of masses $M\lesssim 8M_{\odot}$, after burning up through nuclear processes. most hydrogen and helium that are contained in stellar mass are transformed into carbon, silicon, oxygen or may be iron.
In such stars the most of the mass density is contributed by a non-degenerate ions gas.
So the he internal temperature of most white dwarfs is of order $10^6$ to $10^7 K$. For electrons the Fermi energy $E_f > m_e=0.511 ~MeV \sim ~6 \times 10^9 K$.
Thus the Fermi energy of the electrons in such systems is higher than the kinetic temperature of the environment so it is satisfying the degeneracy condition, so the electrons can be taken to be a zero- temperature Fermi gas.
Because the white dwarfs actually have a non-zero temperature, they have a finite luminosity due to the radiation of heat energy.
The interior of the white dwarf is completely degenerate and the conductivity of the interior is high so it may be considered isothermal.
Cooling of white dwarf occurs only over the surface of the star which has a lower temperature due to radiation.
So that the equation of state for degenerate matter will be inapplicable near the surface of white dwarf.
It means that the surface layer of the white dwarf needs to be treated separately.
The effect of this mechanism in white dwarf is out of the scope of this article and we will consider that the pressure all over the star due to electrons in the degenerate regime with zero-temperature.
According to the uncertainty principle the quantum pressure of degenerate electrons will holding up a white dwarf from Newtonian gravitational collapse. So in order to calculate the thermodynamic properties for that quantum system, we should deal with the degenerate interior of these stars.  Using the deformed density of states, Eq.(\ref{v14}), we can calculate the number density of degenerate electrons as \cite{43}
\begin{equation}\label{c1}
n_e=\dfrac{8\pi}{h^3}\int_{0}^{p_f} \dfrac{p^2dp}{(1-\a p)^4}\simeq
\dfrac{8\pi}{h^3} \le(\dfrac{1}{3} p_f^3 +\a p_f^4\ri)
\end{equation}
Solving Fermi momentum, keeping the terms of order $\sim \a$, one gets
\begin{equation}\label{c2}
p_f=\le( \frac{3h^3n_e}{8\pi} \ri)^{1/3}-\a \le( \frac{3h^3n_e}{8\pi}\ri)^{2/3}
\end{equation}
The pressure of fermions will be calculated using Fermi-Dirac statistics with modified phase space
\begin{equation}\label{c3}
P_0=\dfrac{8\pi}{h^3}\dfrac{1}{\b}\int ln\le[1+z e^{-\b(E-mc^2)}\ri]\dfrac{p^2dp}{(1-\a p)^4}
\end{equation}
where $E^2=c^2p^2+m^2c^4$. Integrate the above integral by parts and then apply the condition of degeneracy; the pressure of electrons in a completely degenerate state will take the form
\begin{eqnarray}\label{c5}
P_0=\dfrac{8\pi}{3h^3}\int^{p_f}_{0} \dfrac{\pa E}{\pa p}\dfrac{p^3dp}{(1-\a p)^3}=\dfrac{8\pi c^2}{3h^3}\int^{p_f}_{0}
\dfrac{1}{\le(c^2p^2+m^2c^4\ri)^{1/2}}\dfrac{p^4dp}{(1-\a p)^3}
\end{eqnarray}
Using the substitution $p=mc\sinh{x}$; one can find that $\frac{p^4dp}{(1-\a p)^3}=m^5c^5\cosh{x}\sinh^4{x}(1+3\a mc\sinh{x})dx$, then
\begin{eqnarray}\label{c8}
P_0=\dfrac{8\pi m^4c^5}{3h^3} \int^{x_f}_{0}\le(1+3\a mc\sinh{x}\ri)\sinh^4{x}dx =\dfrac{8\pi m^4c^5}{3h^3} \le[A(y)+3\a mc B(y)\ri]
\end{eqnarray}
where
\begin{eqnarray}\label{c10}
\nonumber y=\dfrac{p_f}{mc}=\sinh{x_f}~~~~~~~~~~~~~~~~~~~~~~~~~~~~~~~~~~~~~~~~~~~\\
=\le(\frac{3}{\pi}\ri)^{1/3}\frac{h}{2mc}n_e^{1/3}-\a \le(\frac{3}{\pi}\ri)^{2/3}\frac{h^2}{4mc}n_e^{2/3}~~~~~~~~~\\
A(y)=\dfrac{1}{8}\le[3~ln\le(y+\sqrt {1+y^2}\ri)+y\le(2y^2-3\ri)\sqrt{1+y^2} \ri]\\
B(y)=\dfrac{\sqrt{1+y^2}}{15}\le[ 8-4y^2+3y^4\ri]-\dfrac{8}{15}~~~~~~~~~~~~~~~~~~~~~
\end{eqnarray}
For the construction of stellar models we need the matter density $\r$ instead of the
electron number density $n_e$, they are related through the relation $n_e=\frac{\r}{m_uu_e}$, where $u_e$ is called the molecular weight per electron and $m_u=1.6605\times 10^{-27}Kg$ is the atomic mass unit.

When the electron gas is in a low density $y\ll 1$, we can use a non-relativistic dynamics, the pressure in this case looks like
\begin{equation}\label{c14}
P=N_1\r^{5/3}+\a N_2\r^2
\end{equation}
\begin{equation}\label{c15}
N_1=\frac{1}{20}\le(\frac{3}{\pi}\ri)^{2/3}\frac{h^2}{m(m_uu_e)^{5/3}}~~~~~,~~~~~
N_2=-\frac{1}{16} \le(\frac{3}{\pi}\ri)\frac{h^3}{ m(m_uu_e)^2}
\end{equation}
When the electron gas is in a high density $y\gg 1$, that relativistic effect comes to play strongly such that the pressure will look like
\begin{equation}\label{c16}
P=K_1\r^{4/3}+\a K_2\r^{5/3}
\end{equation}
\begin{equation}\label{c17}
K_1=\frac{1}{8}\le(\frac{3}{\pi}\ri)^{1/3}\frac{ch}{(m_uu_e)^{4/3}}~~~~~,~~~~~
K_2=-\frac{1}{10}\le(\frac{3}{\pi}\ri)^{2/3}\frac{ch^2}{(m_uu_e)^{5/3}}
\end{equation}
From these expressions we see that pressure of a completely degenerate electrons depends on density. As the density increases, degeneracy pressure increases as well, and the pressure gradients which develops inside the star is sufficient to support the equilibrium against gravitational contraction. In non-relativistic and extreme relativistic cases the degenerate gas of electrons behaves as a perfect gas with a polytropic equation of state and the pressure decreases due to the presence of quantum gravity.

It is worth mentioning that there are some physical aspects that may alter the equation of state .
It should be corrected because of Coulomb corrections.
At low densities the ions can not be treated as an ideal gas, the Coulomb interaction makes them form a solid lattice structure.
This effect is small, and can be ignored, in the case of astrophysical objects with extreme relativistic limit. It is of order $\sim 10^{-2}$.
This effect should be taken into account in case of low densities with non-relativistic limits which is not the topic of this paper.
In high densities there are a several effects that alter the equation of state.
First, when the density increases, the Fermi energy of the electrons will become a high enough to induce the inverse beta decay reaction.
This process reduces the number of the electrons and thus the electron pressure and the system becomes unstable.
The second process that affect the equation of state is the possibility that nuclear reaction can take place in the crystal lattice at sufficiently high densities.
This arises because of the zero point oscillations of the nuclei which allow them to tunnel through the potential barrier to the neighbouring site and introduce nuclear reaction.
These effects can alter the equation of state at high densities and can be considered a major source of instability in that regime.
Rotation or magnetic field can also affect the structure of white dwarfs.
The effect of these phenomena is not very important for non-relativistic configuration.
But in relativistic limit the radius of the star can increase significantly even with small magnetic potential.
Although this result is of some theoretical significance, there is no evidence for strong magnetic fields in white dwarfs.
Also the mass radius relation is affected by rotation in relativistic regime such that the maximum mass of a rapidly rotating white dwarf can be nearly twice as large as the non-rotating one.
These phenomena are not the topic of this paper.
In the other hand our analysis is based on Newtonian gravity.
At high densities we should take into account the general relativistic corrections which may lead to instability.
This because in Newtonian gravity we balance the gravitational field of the mass by the pressure, so we can arrange the the equation of state such that the pressure is arbitrary high for any given density, thereby balancing any given gravitational force.
In general relativity the pressure contributes an effective mass, and hence increasing the pressure will increase the gravitational force.
Therefore we cannot ensure the system is stable by increasing the pressure arbitrarily.
The effect of quantum gravity in the stability of white dwarf with relativistic correction will be postpone in another research.
In this article we interest in the effect of quantum gravity in mass radius relation for white dwarfs as a tool to determine or put a boundaries on the quantum gravity parameters.

\section{Structure equations for white dwarf}

The evolution of a star may be perceived as a quasi-static process, in which the composition changes slowly, allowing the star to maintain hydrostatic equilibrium and thermal equilibrium. The static structure of a star is obtained from the solution of the set of differential equations known as the stellar structure equations, namely, hydrostatic equilibrium equations, continuity equation, radiative transfer equations and the thermal equilibrium equations \cite{200}. If the mass density $\r$ of the star does not depend on temperature but pressure, i.e. $\r=\r(P)$, we can construct a system of only two equations, namely hydrostatic equilibrium equations for $P$ and $\f$ which can be solved without the other structure equations. $\f$ is the gravitational potential which describes the gravitational field inside the star and is related to the density through Poisson equation. For spherical symmetry and nonrotating stars with uniform composition Poisson equation reduces to
\begin{equation}\label{a1}
\frac{1}{r^2}\frac{d}{dr}\le(r^2\frac{d\f}{dr}\ri)=4\pi G\r
\end{equation}
The time independent solution will be considered. For hydrostatic equilibrium \cite{200}
\begin{equation}\label{a2}
\frac{dP}{dr}=-\frac{d\f}{dr}\r
\end{equation}
According to eqs. (\ref{c14},\ref{c16}), let us assume a general relation between $P$ and $\r$ to take the form
\begin{equation}\label{a3}
P=K_1\r^{\c_1}+\a K_2\r^{\c_2}
\end{equation}
Use Eq.(\ref{a3}) into Eq.(\ref{a2}) and integrate with boundary conditions $\r=0$ and $\f=0$ at the surface of stellar object and using the polytropic index $n=\frac{1}{\c-1}$, one gets
\begin{equation}\label{a4}
\f=-K_1(1+n_1)\r^{1/n_1}-\a K_2(1+n_2)\r^{1/n_2}
\end{equation}
This equation can be solved for $\r$ to get
\begin{equation}\label{a5}
\r=\le[\frac{-\f}{K_1(1+n_1)}\ri]^{n_1}+\a n_1\frac{K_2(1+n_2)}{K_1(1+n_1)} \le[\frac{-\f}{K_1(1+n_1)}\ri]^{m}~~~,~~~m=\frac{n_1}{n_2}+n_1-1
\end{equation}
Define the new dimensionless variables $\xi,w$ by \cite{201}
\begin{equation}\label{a6}
\xi=Ar~~~,~~~A^2=\frac{4\pi G}{\le[K_1(1+n_1)\ri]^{n_1}}(-\f_c)^{n_1-1}~~~,~~~w=\frac{\f}{\f_c}
\end{equation}
Where the index $c$ refers to the center. It is clear that at the center where $r=0$ we have $\xi=0$, $\f=\f_c$, $\r=\r_c$ and therefore $w=1$. We can prove that
\begin{equation}\label{a7}
4\pi G\r=-A^2\f_c\le(w^{n_1}+\a \eta w^m\ri)
\end{equation}
\begin{equation}\label{a8}
\eta=n_1K_2(1+n_2)\frac{1}{\f_c}\le[\frac{-\f_c}{K_1(1+n_1)}\ri]^{n_1/n_2}
\end{equation}
and Eq.(\ref{a1}), with these non-dimensional variable, will take the form
\begin{equation}\label{a9}
\frac{1}{\xi^2}\frac{d}{d\xi}\le(\xi^2\frac{dw}{d\xi}\ri)=-\le(w^{n_1}+\a \eta w^m\ri)
\end{equation}
This is a modified Lane-Emden equation, if we set $\a=0$ we will back to original one. We are only interested in solutions that are finite at the center, $\xi=0$. Equation (\ref{a9}) shows that we have to require $\le(\frac{dw}{d\xi}\ri)_{\xi=0}=0$. Also at the surface of the star where $r=R$ the density vanishes $\r=0$, $\f=0$, $w=0$ and $\xi=\xi_R$ where $\xi_R$ is the value of $\xi$ corresponding to the radius $R$. The mass of the stellar object can be determined as a function of $r$ using the relation $m(r)=\int_{0}^{r}4\pi\r r^2dr$, changing the variable to $\xi$
\begin{eqnarray}\label{a10}
m(\xi)=\frac{1}{A^3}\int_{0}^{\xi}4\pi\r \xi^2d\xi
\end{eqnarray}
This is the mass equation as a function of a dimensionless parameter $\xi$. Using Eqs.(\ref{a7},\ref{a9}) we can prove that $4\pi \r \xi^2d\xi=\frac{A^2\f_c}{G}d\le(\xi^2\frac{dw}{d\xi}\ri)$, the constant $\frac{A^2\f_c}{G}$ can be obtained from Eqs, (\ref{a6}), where $\frac{A^2\f_c}{G}=-4\pi\le[\frac{-\f_c}{k_1(1+n_1)}\ri]^{n_1}$. Use all these in Eq. (\ref{a10}) one gets
\begin{eqnarray}\label{a11}
m(\xi)=4\pi \le[\frac{-\f_c}{K_1(1+n_1)}\ri]^{n_1}\frac{1}{A^3}  \le(- \xi^2 \frac{dw}{d\xi}\ri)
\end{eqnarray}
Eq.(\ref{a4}) is valid for the central potential and corresponding density, so we have the equation
\begin{equation}\label{a12}
\f_c=-K_1(1+n_1)\r_c^{1/n_1}-\a K_2(1+n_2)\r_c^{1/n_2}
\end{equation}
In fact Eq.(\ref{a6}) shows that $\frac{\xi}{r}$ is always the constant $A$ which can be fixed though the relation $A=\frac{\xi_R}{R}$. In turn the total mass of the star can be determinted from Eq. (\ref{a11})
\begin{equation}\label{a14}
M=4\pi \le[\frac{-\f_c}{K_1(1+n_1)}\ri]^{n_1} R^3 \le(- \frac{1}{\xi} \frac{dw}{d\xi}\ri)_{\xi=\xi_R}
\end{equation}
If we introduce the mean density as $\bar{\r}=\frac{3M}{4\pi R^3}$
\begin{equation}\label{a15}
\bar{\r}=3 \le[\frac{-\f_c}{K_1(1+n_1)}\ri]^{n_1} \le(- \frac{1}{\xi} \frac{dw}{d\xi}\ri)_{\xi=\xi_R}
\end{equation}
From Eq.(\ref{a6}) we can write
\begin{equation}\label{b1}
\frac{1}{A^2}=\le(\frac{r}{\xi}\ri)^2=\frac{\le[K_1(1+n_1)\ri]^{n_1}}{4\pi G}(-\f_c)^{1-n_1}
\end{equation}
From this equation we can obtain the radius of the star using the replacement $(r,\xi)\rightarrow(R,\xi_R)$
\begin{equation}\label{b2}
R=\le[-\frac{K_1(1+n_1)}{\f_c}\ri]^{\frac{n_1}{2}}\le(-\frac{\f_c}{4\pi G}\ri)^{1/2}\xi_R
\end{equation}
Use Eq.(\ref{b2}) into Eq.(\ref{a14}) to eliminate $R$
\begin{equation}\label{b3}
M=4\pi\le[-\frac{K_1(1+n_1)}{\f_c}\ri]^{\frac{n_1}{2}}\le(-\frac{\f_c}{4\pi G}\ri)^{3/2}\le(- \xi^2\frac{dw}{d\xi}\ri)_{\xi=\xi_R}
\end{equation}
We can eliminate $\f_c$ between Eqs.(\ref{b2},\ref{b3}), one gets $M \thicksim R^\frac{3-n_1}{1-n_1}$ which shows that the presence of quantum gravity does not change the proportionality relation between mass and radius of the star put changes only the proportionality constant. for non-relativistic case where $n_1=3/2$ gives $M \thicksim R^{-3}$, in words the star has a smaller radius for a larger value of mass. for ultra-relativistic case where $n_1=3$  mass M becomes independent of radius and the value of mass in this case Chandrasekhar mass. If we let $\a\rightarrow 0$ the mass and radius limits will go back to original expression \cite{200}, namely

\begin{equation}\label{b4}
R=\le[\frac{K_1(1+n_1)}{4\pi G}\ri]^{1/2}\r_c^{\frac{1-n_1}{2n_1}}\xi_R~~~~~,~~~~~M=4\pi\le[\frac{K_1(1+n_1)}{4\pi G}\ri]^{3/2}\r_c^{\frac{3-n_1}{2n_1}}\le(-\xi^2\frac{dw}{d\xi}\ri)_{\xi=\xi_R}
\end{equation}

\section{Numerical Solutions of the Modified Lane-Emden Equations}

In this article we are interested only with Chandrasekhar limits, so we will consider only the degenerate pressure in ultra-relativistic regime where $n_1=3$ and $n_2=\frac{3}{2}$. In this case, it is clear that for $\a=0$, Eq.(\ref{a9}) goes back to usual Lane-Emden equation, also the mass of the star Eq.(\ref{b3}) and central density are no longer coupled which means that the central density can be arbitrary and therefor also the radius. In this case we got the Chandrasekhar mass limit $M_{Ch}=1.46M_{\odot}$, for $u_e=2$. Indeed no white dwarf has been found which exceeds this mass\cite{201,202,203,204}.

\textbf{The existence of} quantum gravity leads the presence of central density dependence. The solution from modified Lane-Emden equation depends on central density and in turn will affect the mass and radius of the star. So we need a numerical value for central density, all white dwarfs central density ranged from $10^8$ to $10^{12} Kg/m^3$ \cite{301,302}, so we will use three values namely $\r_c=10^8, 10^{10}$ and $10^{12} Kg/m^3$.

In \cite{42} The mass-radius relation for white dwarfs, with using constant density, is studied under the effect of quantum gravity. Reduction in degenerate pressure leads to more control of gravity collapse and hence a contraction in the radius of the star. According to this study the modified star radius is
\begin{eqnarray}\label{d1}
 R \approx R_{Ch} \le( 1-\dfrac{1}{5}\a m_ec  \dfrac{3\tau+4}{\tau^{3/2}} \ri)
\label{e13}
\end{eqnarray}
Where $\t=1-\le(\frac{M}{M_{Ch}}\ri)^{2/3}$ and $R_{Ch}$ is Chandrasekhar radius limit. The remarkable aspect in this relation is that the quantum gravity clearly shows its influence for $M\rightarrow M_{Ch}$. It means that the quantum gravity reduces the star radius and the reduction become significant when the mass of the star gets very close to the Chandrasekhar limit in high density regime. The author used the upper bound of $\a_0$ (i.e. $\a_0\approx 10^{17}$). Masses of two white dwarfs, namely Wolf 485 A and G 156-64 \cite{304}, are used to estimate the star radius correction which leads to $\triangle R\simeq 52~m$. This modification  does not exceed  the desired correction. The correction should be of order $\sim 10^6~m$.

We can estimate the suitable range of quantum gravity parameter $\a_0$ which will be used by a comparison between the experimental values of mentioned stars (mass and radius) and the theoretical values predicted by Eq.(\ref{d1}). These calculations will lead to $\a_0 \approx 10^{21}$. In fact this value is  inconsistent with experimental results which give a length scale bigger than electroweak length. But this value goes in parallel with the results that are predicted in \cite{26}. "With more accurate measurements in the future, this bound is expected to get reduced by several orders of magnitude, in which case, it could signal a new and intermediate length scale between the electroweak and the Planck scale" \cite{26,306}.

Recently, Ali at al \cite{308} calculated the modified Schwarzshild metric with quantum gravity to find corrections to some relativistic phenomena namely deflection of light, time delay of light, perihelion precession, and gravitational redshift. After that they compere the final results with experiment to set upper bounds on the GUP parameter $\a_0$. The value of $\a_0$ lies in between $10^{35}$ and $10^{41}$. These values are very large in comparison with those predicted from quantum mechanical systems. "However, investigating the implications of the GUP on gravitational phenomena might prove useful for understanding the effects of quantum gravity in that regime. In addition, because the GUP is model independent, this understanding can help to evaluate the results of different theories of quantum gravity."\cite{308}. keeping in mind all these investigations, with satisfaction, we can use the values $\a_0=10^{17}, 10^{19}$ and $10^{21}$ although it leads to a length scale bigger than electroweak length. We expect that this range of gravity parameter will reduce Chandrasekhar limits.

Values of mentioned $\a_0$ and $\r_c$ can be used with the boundary conditions $w(0)=1$, $w'(0)=0$ and $u_e=2$ (for He, C, O, ...) in solving Eq.(\ref{a9}), the values of $\xi_R$ and the derivative $\frac{dw}{d\xi}$ can be obtained when $w(\xi)=0$. This result should be used in determining the radius and mass limits, Eqs.(\ref{b2},\ref{b3}). The results are listed as follow
\begin{equation}
\begin{array}{ccccc}
\nonumber
\a_0~~~~~     & \r_c~~~~~&  \xi_R~~~~    & M_{Ch}/M_{\odot}~~~~~& R_{Ch}/R_{\odot} \\
            ~\\
10^{17}~~~~~  & 10^{8}~~~~~ & 6.897~~~~  & 1.4564~~~~~&0.10365 \\
            & 10^{10}~~~~~& 6.897~~~~  & 1.4564~~~~~&0.02233 \\
            & 10^{12}~~~~~& 6.897~~~~  & 1.4563~~~~~&0.00481 \\
            ~\\
10^{19}~~~~~  & 10^{8}~~~~~~&  6.896~~~~  & 1.4562~~~~~&0.10365 \\
            & 10^{10}~~~~~&  6.895~~~~ & 1.4557~~~~~&0.02234 \\
            & 10^{12}~~~~~&  6.888~~~~ & 1.4535~~~~~&0.00482 \\
            ~\\
10^{21}~~~~~  & 10^{8}~~~~~~&  6.858~~~~  & 1.4430~~~~~&0.10469 \\
            & 10^{10}~~~~~&  6.732~~~~ & 1.3983~~~~~&0.02349 \\
            & 10^{12}~~~~~&  6.407~~~~ & 1.2499~~~~~&0.00672 \\
\end{array}
\end{equation}
Some facts can be drawn from the previous numerical results. There facts can be summarized as follows:

1-~ the quantum gravity leads to a decrease in star degenerate pressure which in turn supports star collapse and a reduction in star radius which is proportional to the quantum gravity parameter. This result is reported in \cite{42} where the density of the star is constant. Introduction of realistic density inverts this behavior. We can draw from the above numerical solution for constant central density the radius of the star increases with increasing gravity parameter $\a_0$.

2-~ Numerical solutions show that the radius of the star decreases with increase in central density. These results are consistent with the current observation which indicates that white dwarfs have smaller radii than theoretical predictions \cite{304,309}. This means that quantum gravity and central density can be a reasonable tool to explain the smallness in radii observation.

3-~ For $\a_0=0$ the mass limit is $M_{Ch}/M_{\odot}=1.4564$, the results show that the Chandrasekhar mass limit decreases as the central density and quantum gravity parameter $\a_0$ increases. Also the change in mass limit is significant with large gravity parameter.

\section{Conclusion}
This study reviewed in brief the most important quantum gravity approaches. The effect of generalized uncertainty principle in white dwarf physics was investigated to find the maximum value of the radius and mass of these stars. This type of generalized uncertainty principle is suitable because it leads to a reduction in mass and star radius which is consistent with current observations. This effect is what makes the central density come to play in mass formula. This way there are two new parameters, namely, quantum gravity and central density together that can be help to make the theoretical values corresponding to experimental observations apply in a better manner. In this work we treat the star as completely degenerate, but the reality is the white dwarf is so dense that matter becomes degenerate in its interior. At the surface of a white dwarf the density is that lower and no degeneracy will be found in its outer layers. For finite temperature not all the electrons will be densely packed in momentum space in the cells of lowest possible momentum. If the temperature is sufficiently high we expect them to have a Boltzman distribution. But the quantum gravity does not make a change in classical gases \cite{310} so the pressure of the star should be corrected for these reasons. In future these points should be taken into account. And to form a complete picture of the white dwarf also we should take into account the chemical composition of the star and its complete degenerate pressure formula in order to unify or determine a suitable set of parameters that are compatible with all experimental data. Although we used large values of the gravity parameter $\a_0$ which are incompatible with theoretical conditions, the most recent theoretical studies in astro and space physics confirm that these values need to be large in order to follow the experimental observations. Perhaps there are draw backs in the theory of quantum gravity that need to be avoided in the future.

\section{Acknowledgments}
This research is supported by Benha university (www.bu.edu.eg/).

\end{document}